\documentclass[aps,prd,twocolumn,superscriptaddress,floatfix,showkeys,showpacs]{revtex4}
\usepackage{pdfsync}
\usepackage{mathptmx}
\usepackage[utf8]{inputenc}
\usepackage[T1]{fontenc}
\usepackage{slashed}
\usepackage{times}
\usepackage{amssymb,amsfonts,amsmath,amsthm}
\usepackage{dsfont,bbm}
\usepackage{dcolumn}
\usepackage{epsf}
\usepackage{graphicx}
\usepackage[caption=false]{subfig}
\usepackage{dsfont}
\usepackage{slashed}
\usepackage[active]{srcltx}
\usepackage[usenames]{color}

\begin{document}

\title{Drell-Yan-like processes and duality}

\author{I.~V.~Anikin}
\email{anikin@theor.jinr.ru}
\affiliation{Bogoliubov Laboratory of Theoretical Physics, JINR,
             141980 Dubna, Russia}
\author{L.~Szymanowski}
\affiliation{National Centre for Nuclear Research (NCBJ),
            00-999 Warsaw, Poland}
\author{O.~V.~Teryaev}
\affiliation{Bogoliubov Laboratory of Theoretical Physics, JINR,
             141980 Dubna, Russia}
\author{N.~Volchanskiy}
\affiliation{Bogoliubov Laboratory of Theoretical Physics, JINR,
             141980 Dubna, Russia}
\affiliation{Research Institute of Physics, Southern Federal University,
             344090 Rostov-on-Don, Russia}

\begin{abstract}
We calculate the gauge invariant Drell-Yan-like hadron tensors.
In connection with new COMPASS results, we predict the new single spin asymmetry which probes gluon poles together with
chiral-odd and time-odd functions.
The relevant pion production as a particular case of Drell-Yan-like process has been discussed.
For the meson-induced Drell-Yan process, we model an analog of the twist three distribution function,
which is a collinear function in inclusive channel,
by means of two non-collinear distribution amplitudes which are associated with
exclusive channel. This modelling demonstrates the fundamental duality between different factorization regimes.

\end{abstract}
\date{\today}
\maketitle


The current studies of hadron structure involve both semi-inclusive and exclusive processes.
They are described by Transverse Momentum Dependent (TMDs) and Generalized Parton Distributions (GPDs), respectively.
The transitions, duality and matching between these regimes are of much importance for the coherent QCD description of hadron structure.
Here we concentrate on the manifestation of such effects in the pion-nucleon Drell-Yan process at large $x_F$, when
pion is described by wave functions and distribution amplitudes rather than parton distributions.

It has been shown long ago that the specific effects related to the high-twist corrections
lead to the sizeable non-scaling and non-factorizing contributions to the unpolarized cross sections
for the Drell-Yan-like processes in the well-defined kinematic regimes of
a large fraction $x$ \cite{Berger:1979du, Berger:1979xz}.
Also, the inclusive production of dimuons from the
hard scattering of pions on an unpolarized nuclear target and the similar process with
longitudinally polarized protons have been analysed in \cite{Bakulev:2007ej}. In both cases, it has demonstrated
the role played by the pion bound state in terms of the pion distribution amplitudes.
Moreover, in \cite{Brandenburg:1994wf} it is shown that
the angular distribution is rather sensitive to the shape of the pion distribution amplitude
in the kinematic region where one of the pion constituents is off-shell.
In the kinematic regime where the photon has a large longitudinal
momentum fraction, the cross-section and the single
spin asymmetry for the dimuon production with taking into account pion bound state
effects are calculated in \cite{Brandenburg:1995pk}.
The predictions of \cite{Brandenburg:1995pk} are directly proportional to the pion distribution
amplitude. Therefore, the measurement of the polarized Drell-Yan cross section
can determine the shape of the pion distribution amplitude.

To the present day the study of hadron (in particular nucleon) composite structure
is the most important subject of hadron physics.
From the experimental viewpoint, one of the wide-spread and useful instruments for such investigations is
the single spin asymmetry (SSA). Especially, the single transverse spin asymmetry
opens the access to the three-dimensional nucleon structure owing to
the nontrivial connection between the transverse spin and the parton transverse momentum dependence
(see, for example, \cite{Angeles-Martinez:2015sea, Boer:2011fh, Boer:2003cm, Kang:2011hk, Boer:2011fx, Arnold:2008kf}).
There are several experimental programs which pursue the measurements with Drell-Yan-like processes,
RHIC \cite{Bai:2013plv}, COMPASS \cite{Baum:1996yv, COMPASS} and future NICA \cite{Kouznetsov:2017bip, Savin:2016arw}.

In the paper, we study the Drell-Yan and pion production hadron tensors related
to the meson-baryon processes with the essential
spin transversity and ``primordial'' transverse momenta.
The main attention has been paid for the gluon poles which are manifested in the
corresponding distribution functions or/and amplitudes.

Inspired by the recent experimental studies by COMPASS \cite{COMPASS},
we propose an approach to calculate the gauge invariant meson-induced Drell-Yan hadron tensor which
finally gives a prediction for the single transverse spin asymmetry.
We focus on the case where one of pion (or meson) distribution amplitudes has been projected onto the chiral-odd combination.
In turn, the pion chiral-odd distribution amplitudes also separate the chiral-odd tensor combination in the nucleon matrix element.
The access to the single spin asymmetries, in particular the angular dependence, induced by
chiral-odd  and time-odd distribution functions/amplitudes has been opened only owing to the gluon pole presence.
In other words, the angular dependence of SSA predicted in the paper
can give implicit evidence for the gluon pole observation in COMPASS experiments.

Moreover, even in the collinear collision of hadron beams it is possible to get experimentally
some evidence for the leading role of the transverse parton movements inside hadrons.
Indeed, thanks for the frame-independency, within the so-called Collins-Soper frame
we can study the angular dependence of SSA as a function of $\varphi\sim \vec{\bf S}_\perp\wedge\vec{\bf P}_{\perp}$
provided the gluon poles presence.
The nontrivial angular dependence of SSA can be treated as a signal for the transverse ``primordial'' momentum dependence.
Thus, we impose the single spin asymmetries which are reachable in COMPASS and can probe simultaneously gluon poles,
duality, chiral-odd and time-odd functions.


Further, let us go over to kinematics,
we study the meson-induced Drell-Yan process
and semi-inclusive deep-inelastic scattering with pion production
where baryons are transversely polarized:
$M(P_1) + N^{(\uparrow\downarrow)}(P_2) \to \gamma^*(q)+ \bar q(K) + X(P_X)
\to\bar q(K) + \ell(l_1)+\bar\ell(l_2) + X(P_X)$
and $N^{(\uparrow\downarrow)}(P_2) + \ell(l_1)
\to M(P_1)+ \ell(l_2) + \bar q(K) + X(P_X)$.
The virtual photon producing the lepton pair ($l_1+l_2=q$) has a large mass squared
($q^2=Q^2$) while the transverse momenta are small and integrated out.

The Sudakov decompositions take the forms
(for the sake of shortness, we omit the four-dimension indices)
\begin{eqnarray}
\label{Sudakov-decom-1}
&&P_1\approx \frac{Q}{x_B \sqrt{2}}\, n^*  + P_{1\,\perp} \, ,
P_2\approx \frac{Q}{y_B \sqrt{2}}\, n  + P_{2\,\perp} \, ,
\\
&&S\approx \frac{\lambda}{M_2}\,P_2 + S_\perp\,
\end{eqnarray}
for the hadron momenta and spin vector;
\begin{eqnarray}
\label{Sudakov-decom-2}
q=\frac{Q}{\sqrt{2}}\, n^* +\frac{Q}{\sqrt{2}}\,n + q_{\perp},\, \quad q_\perp^2\ll Q^2\, ,
\end{eqnarray}
for the photon momentum.
The hadron momenta $P_1$ and $P_2$ have the plus and minus dominant light-cone
components, respectively. Accordingly, the dominant quark and gluon momenta $k_1$ and $\ell$ lie
along the plus direction while the dominant antiquark momentum $k_2$ -- along the minus direction.

We also define the Collins-Soper frame as \cite{Barone:2001sp}
\begin{eqnarray}
\label{CS-frame}
&&\widehat{t}^\mu=\frac{q^\mu}{Q},\quad \widehat{x}^\mu=\frac{q_{\perp}^{\mu}}{Q_\perp},\quad
\widehat{z}^\mu  = \frac{x_B}{Q} \widetilde{P}_1^\mu - \frac{y_B}{Q} \widetilde{P}_2^\mu,
\end{eqnarray}
where $\widetilde{P}_1= {P}_1 - q/(2x_B)$ and $\widetilde{P}_2 = {P}_2 - q/(2y_B)$.
In what follows we are not so precise about writing the covariant and contravariant vectors in any kinds of
definitions and summations over the four-dimensional vectors, except the cases where this notation may lead to misunderstanding.
We can also write down that
\begin{eqnarray}
\label{lcb}
\sqrt{2} n^* = \widehat{t} + \widehat{z} - \frac{Q_\perp}{Q} \widehat{x},
\quad
\sqrt{2} n = \widehat{t} - \widehat{z} - \frac{Q_\perp}{Q} \widehat{x}.
\end{eqnarray}
With minor modifications this reference frame is suitable for the direct pion production as well.


For the processes we consider, we deal with a large $Q^2$ and, therefore, we apply the
factorization theorem to get the corresponding hadron tensor factorized in the form of convolution:
\begin{eqnarray}
\label{Fac-DY}
\text{Hadron tensor} = \{\text{Hard (pQCD)}\} \otimes
\{\text{Soft (npQCD)} \}\,.
\end{eqnarray}
Based on DY kind of processes, it is natural to study the role of twist $3$ by exploring
of different kinds of asymmetries. In particular,
the left-right asymmetry which means that the transverse momenta
of the leptons or/and hadrons are correlated with the directions
$\vec{\textbf{S}}_\perp\wedge \widehat{\mathbf{z}}$  and
$\vec{\textbf{S}}_\perp\wedge \widehat{\mathbf{x}}$, see Eqn.~(\ref{CS-frame}).

The single spin asymmetries (SSAs) under our consideration is given by
\begin{eqnarray}
\label{SSA-1}
{\cal A}= (d\sigma^{(\uparrow)} - d\sigma^{(\downarrow)})/(d\sigma^{(\uparrow)} + d\sigma^{(\downarrow)}),
\end{eqnarray}
with (see, \cite{Arnold:2008kf})
\begin{eqnarray}
\label{dsigma}
\frac{d\sigma^{(\uparrow\downarrow)}}{d^4 q d\Omega}=\frac{\alpha^2_{em}}{2j q^4}
{\cal L}_{\mu\nu}\, H_{\mu\nu}\, ,
\end{eqnarray}
where $j$ is the standard flux factor, $d\Omega$ specifies the frame angle orientations.
In Eqn.~(\ref{dsigma}), ${\cal L}_{\mu\nu}$ implies the unpolarized lepton tensor and
$H_{\mu\nu}$ stands for the hadronic tensor.
Since we dwell on the unpolarized lepton case which
leads to the real lepton tensor, the hadron tensor $H_{\mu\nu}$ has to be real one as well.
Moreover, as a rule, the hadron tensor includes at least two non-perturbative blobs
which are associated with two different
dominant (the light-cone plus and minus) directions.

\begin{figure}[ht]
\centerline{
\includegraphics[width=0.35\textwidth]{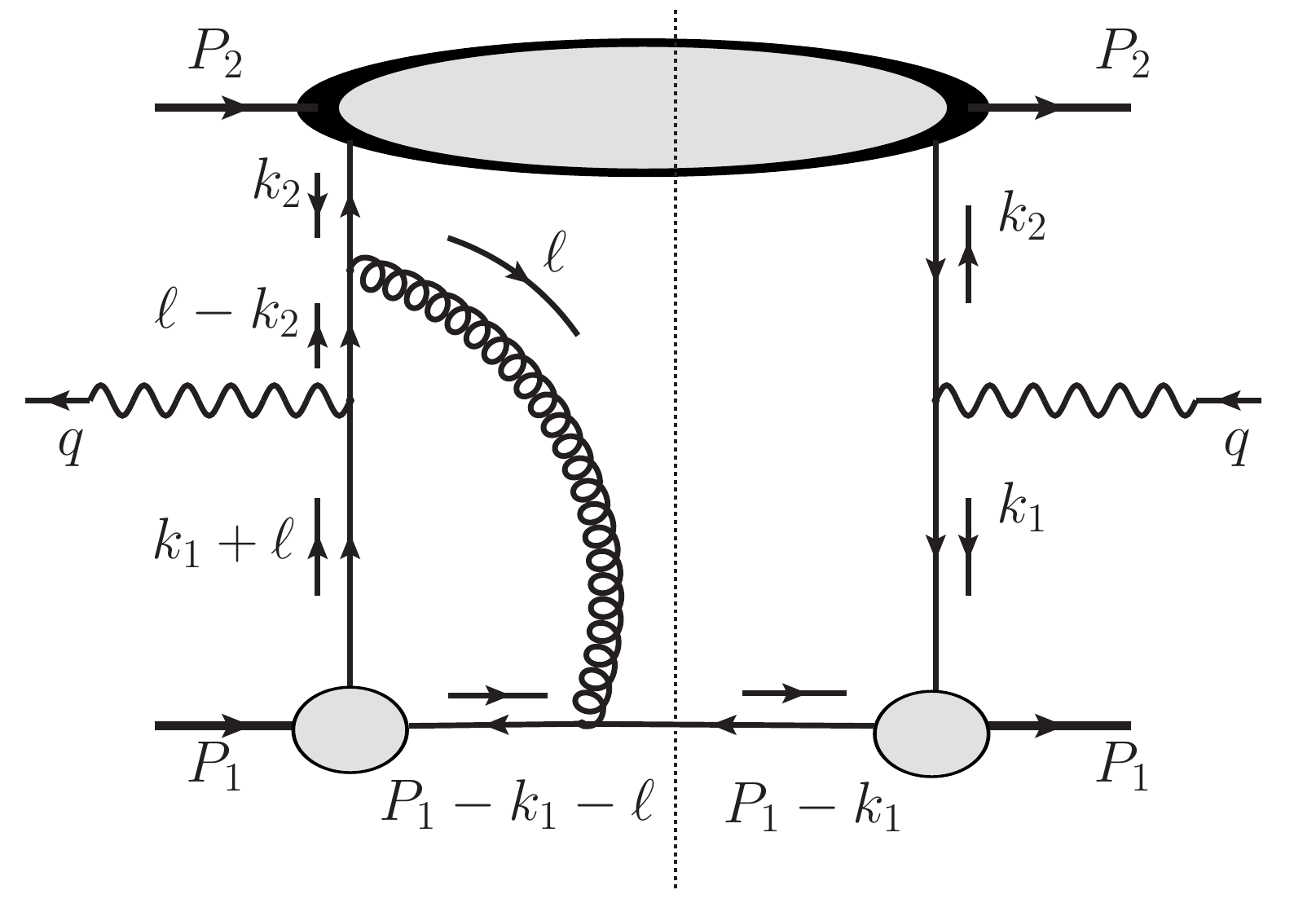}}
  \caption{The Feynman diagrams which contribute to the polarized Drell-Yan hadron tensor:
  the standard diagram.}
\label{Fig-DY-1-2-a}
\end{figure}

\begin{figure}[ht]
\vspace{0.3cm}
\centerline{\includegraphics[width=0.35\textwidth]{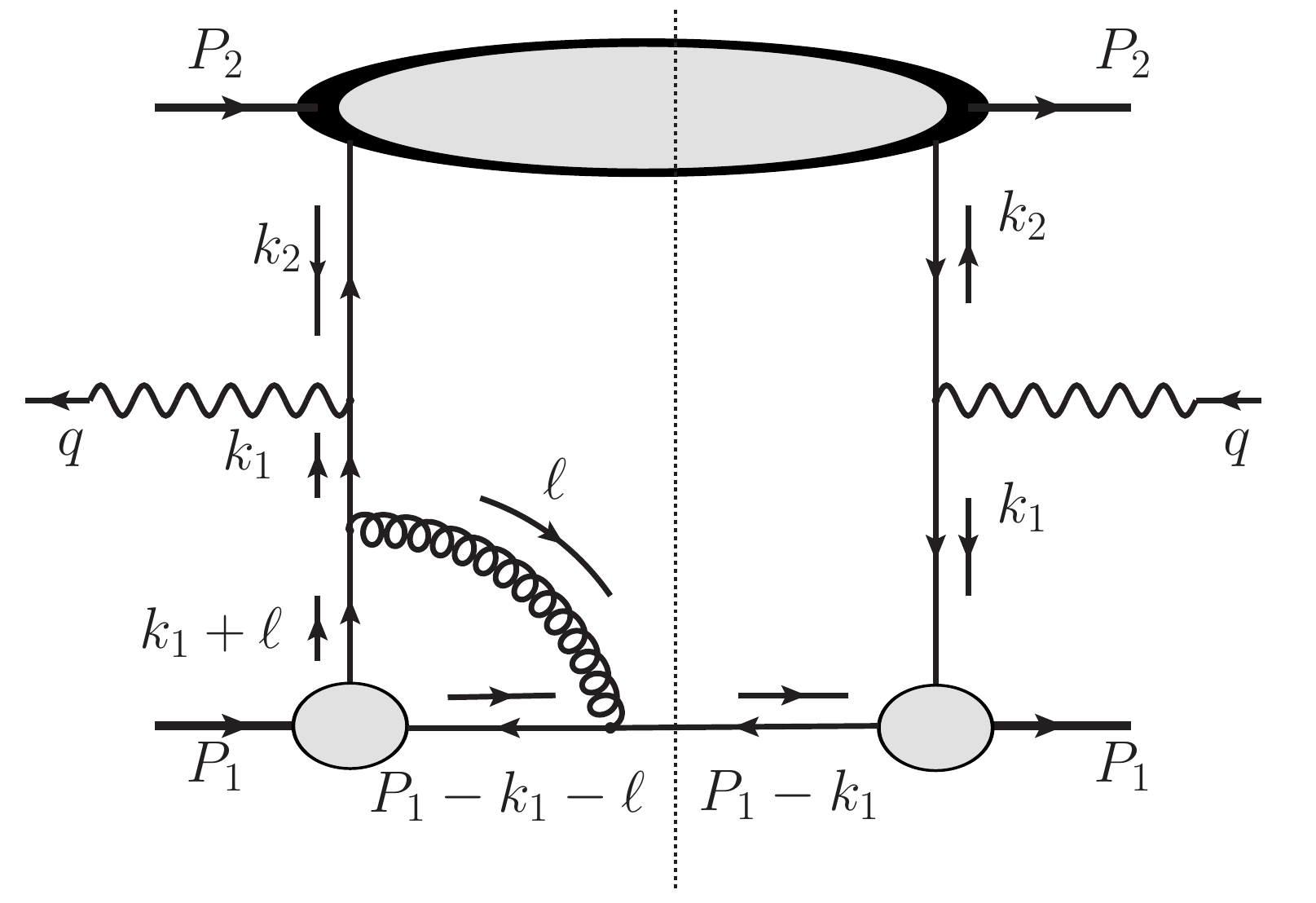}}
  \caption{The Feynman diagrams which contribute to the polarized Drell-Yan hadron tensor:
  the non-standard diagram.}
\label{Fig-DY-1-2-b}
\end{figure}

Following \cite{Anikin:2010wz, Anikin:2015xka, Anikin:2015esa}, we continue to
study the gluon pole influence on the different asymmetries.
We now focus on the cases where the upper non-perturbative blob depicted
in Figs.~\ref{Fig-DY-1-2-a} and \ref{Fig-DY-1-2-b}
corresponds to the matrix elements, first, with the spin transversity
and, second, with the ``primordial'' hadron transverse momentum. The corresponding matrix elements
can be parametrized by either the chiral-odd or time-odd twist two functions, {\it i.e.}
(see below Eqns.~(\ref{Upper-Phi}) and (\ref{Upper-Phi-2}))
\begin{eqnarray}
\label{UpperDY}
&&\langle P_2, S_\perp | \bar\psi\, \sigma^{-\perp} \,
\psi |S_\perp, P_2 \rangle\stackrel{{\cal F}}{\sim} \varepsilon^{- \perp S^{\perp} P_{2}}
\, \bar h_{1}(y),
\nonumber\\
&&\langle P_2, S_\perp | \bar\psi\, \gamma^- \,
\psi |S_\perp, P_2 \rangle\stackrel{{\cal F}}{\sim} i\varepsilon^{- + S_\perp P_{2}^\perp}
\, \bar f_T(y).
\end{eqnarray}
We remind that for the upper blob the dominant light-cone direction is a minus direction.

On the other hand, for the lower non-perturbative blob we replace the twist three $B^V$-function
of \cite{Anikin:2010wz} with two distribution amplitudes with twist two and three,
see Figs.~\ref{Fig-DY-1-2-a} and \ref{Fig-DY-1-2-b}. Schematically, this can be demonstrated as
\begin{eqnarray}
\label{LowerDY}
&&
\langle P_1, S | \big[ \bar\psi\,
A^{\alpha}_\perp \,\psi \big]^{\text{tw-3}}|S, P_1 \rangle
\Big|_{\text{\cite{Anikin:2010wz}}}
\Longrightarrow
\\
&&
{\cal D}^{\alpha\beta} \, \langle 0 | \big[\bar\psi\, \psi \big]^{\text{tw-2}}| S, P_1 \rangle \, \gamma_\beta \,
 \langle P_1, S | \big[\bar\psi\, \psi\big]^{\text{tw-3}} | 0 \rangle
\Big|_{\text{this work}},
\nonumber
\end{eqnarray}
where ${\cal D}^{\alpha\beta}$ stands for the gluon propagator.
Here, all spinor indices in the corresponding matrix element combinations
are open.
Notice that the replacement shown in Eqn.~(\ref{LowerDY})
gives us the possibility to study the so-called gluon poles in the most explicit form.
Indeed, as it is shown below (see, Eqn.~(\ref{GP})), the longitudinal dominant part of gluon propagator finally generates
the gluon pole with the certain complex prescription (arisen from the contour gauge)
which compensates the complexity of (\ref{UpperDY}).


We are now in position to discuss the derivation of hadron tensor.
We begin with the hadron tensor that relates to the standard diagram, see Fig.~\ref{Fig-DY-1-2-a}.
Throughout the paper, we adhere the terminology and
the {\it collinear} factorization procedure as described
in \cite{Anikin:2010wz, Anikin:2015xka, Anikin:2015esa}.
The so-called standard diagram (which exists even if $B^V$-function in \cite{Anikin:2010wz} is real)
implies the diagram depicted in Fig.~\ref{Fig-DY-1-2-a}.

Before factorization, the standard diagram gives the hadron tensor
(all prefactors are included in the integration measures)
\begin{eqnarray}
\label{HadTen-St}
&&{\cal W}_{\mu\nu}^{(\text{stand.})}=\int (d^4k_1)\,(d^4k_2) \, \delta^{(4)}(k_1+k_2-q)
\nonumber\\
&&\times\int (d^4\ell)\, {\cal D}_{\alpha\beta}(\ell)
\text{tr}\big[ \gamma_\nu \Gamma \gamma_\alpha S(\ell-k_2) \gamma_\mu \Gamma_1 \gamma_\beta \Gamma_2
\big]
\nonumber\\
&&\times\bar\Phi^{[\Gamma]}(k_2)\,
\Phi_{(2)}^{[\Gamma_1]}(k_1; \ell)\,
\Phi_{(1)}^{[\Gamma_2]}(k_1)\, \delta\big( (P_1-k_1)^2 \big) \,,
\end{eqnarray}
where
\begin{eqnarray}
\label{barPhi-func}
&&\bar\Phi^{[\Gamma]}(k_2)= \int\hspace{-0.4cm}\sum\nolimits_{X}\,\int (d^4\eta_2)\, e^{-ik_2\eta_2}\,
\\
&&\times\langle P_2, S_\perp |\text{tr}\big[ \psi(0)|P_{X}\rangle \, \langle P_{X} | \bar\psi(\eta_2) \Gamma \big] | S_\perp , P_2\rangle\,
\nonumber
\end{eqnarray}
and
\begin{eqnarray}
\label{Phi2-func}
&&\Phi_{(2)}^{[\Gamma_1]}(k_1; \ell)=
\\
&& \int (d^4\eta_1)\, e^{i(P_1-\ell-k_1)\eta_1}\, \langle 0 | \bar\psi(\eta_1) \,\Gamma_1 \psi(0) | P_1\rangle\,,
\nonumber
\end{eqnarray}
\begin{eqnarray}
\label{Phi1-func}
&&\Phi_{(1)}^{[\Gamma_2]}(k_1)= \int (d^4\xi)\, e^{-ik_1\xi}\,
\langle P_1 | \bar\psi(\xi) \,\Gamma_2 \psi(0) | 0\rangle\,.
\end{eqnarray}
In  Eqn.~(\ref{HadTen-St}), we write explicitly the $\delta$-function which shows that
the quark operator with $P_1-k_1$ corresponds to the on-shell fermion.

For the cases under our consideration,
we choose the $\gamma$-structure in Eqns.~(\ref{HadTen-St}) and (\ref{Phi2-func}), (\ref{Phi1-func}) to be
\begin{eqnarray}
\label{gamma-str}
&&\Gamma\otimes \bar\Phi^{[\Gamma]} \Rightarrow  \gamma^+
\otimes \bar\Phi^{[\gamma^-]} \oplus \sigma^{+\perp}
\otimes \bar\Phi^{[\sigma^{-\perp} ]},
\\
&&\Gamma_1\otimes \Phi_{(2)}^{[\Gamma_1]} \Rightarrow  \gamma^-(\gamma_5) \otimes \Phi_{(2)}^{[\gamma^+(\gamma_5)]},
\nonumber\\
&&\Gamma_2\otimes \Phi_{(1)}^{[\Gamma_2]} \Rightarrow
\gamma_\rho^\perp (\gamma_5)
\otimes \Phi_{(1)}^{[\gamma^\perp_\rho(\gamma_5)]}
 \oplus
 \sigma^{+ -} (\gamma_5)
 \otimes \Phi_{(1)}^{[\sigma^{- +}(\gamma_5) ]}
\nonumber
\end{eqnarray}
which correspond to the nucleon parton distribution of twist two and
the pion or/and rho-meson distribution amplitudes of twist two and three, respectively.
However, the concrete type of hadrons does not play a crucial role in our consideration.

The next item is to perform the factorization procedure for the hadron tensor.
We are not going to stop on all of factorization stages (the comprehensive description of factorization can be found
in many papers, see, for example, \cite{An-ImF, Belitsky:2005qn, Diehl:2003ny, Braun:2011dg}).
Instead, we dwell on the corresponding parton functions parametrizing the non-perturbative matrix elements which
appear after the collinear factorization procedure.
For the twist-$2$ distribution function which characterizes the upper blob, we have
\begin{eqnarray}
\label{Upper-Phi}
&&\bar\Phi^{[\sigma^{-\perp}]}(y)\stackrel{\text{def.}}{=}
\\
&&\int(d^4k_2) \delta(y-k_2^-/P_2^-) \bar\Phi^{[\sigma^{-\perp}]}(k_2) = \varepsilon^{-\perp S^{\perp} P_2}
\bar h_{1}(y),
\nonumber\\
&&\bar\Phi^{[\gamma^-]}(y)\stackrel{\text{def.}}{=}
\\
\label{Upper-Phi-2}
&&\int(d^4k_2) \delta(y-k_2^-/P_2^-) \bar\Phi^{[\gamma^-]}(k_2) = i\varepsilon^{-+S_\perp P_2^\perp} \bar f_T(y).
\nonumber
\end{eqnarray}
The other object is an analog of $B^V$-function expressed through the gluon propagator and functions $\Phi_{(1)}$, $\Phi_{(2)}$.
We have (here, $x_{21}=x_2-x_1$)
\begin{eqnarray}
\label{Lower-Phi}
&&\mathbb{B}^{[\Gamma_2,\,\Gamma_1]}_{\alpha\beta}(x_1,x_2)\stackrel{\text{def.}}{=}
\\
&&\int(d^4k_1) \delta(x_1-k_1^+/P_1^+) \, \int(d^4\ell) \delta(x_{21}-\ell^+/P_1^+)\,
\nonumber\\
&&\times \frac{d_{\alpha\beta}(\ell)}{2\ell^+\ell^- - \vec{\ell}^{\,2}_\perp+i0}\,
\Phi_{(2)}^{[\Gamma_1]}(k_1; \ell)\, \Phi_{(1)}^{[\Gamma_2]}(k_1)\, \delta\big( (P_1-k_1)^2 \big)
\nonumber
\end{eqnarray}
where the gluon propagator has been written in the explicit form with the causal prescription and
$d_{\alpha\beta}(\ell)$ given by the axial (contour) gauge $A^+=0$ (see, \cite{Anikin:2016bor}).

Since we are interested in the gluon pole decoded in Eqn.~(\ref{Lower-Phi}) we have to keep only
the transverse gluon contributions to the gluon propagator. Indeed, the gluon propagator generated
by $\langle A^i_\perp A^-\rangle$ is associated with the essential transverse component of gluon momentum
$\ell^{\,i}_\perp$. This leads to the case where there are no any sources for the gluon pole at $x_1=x_2$ \cite{Anikin:2015esa, Braun}.
On the other hand, $\langle A^i_\perp A^-\rangle$ part of gluon propagator is gauge-dependent and disappears
ultimately in the physical observables and hadron tensor.

Therefore, Eqn.~(\ref{Lower-Phi}) can be rewritten in the following form
\begin{eqnarray}
\label{Lower-Phi-2}
\mathbb{B}_{\alpha\beta}^{[\Gamma_2,\,\Gamma_1]}(x_1,x_2)=
\frac{g^\perp_{\alpha\beta}}{2P_1^+ (x_2-x_1)}
\Phi_{(1)}^{[\Gamma_2]}(x_1)\hspace{-0.2cm}
\stackrel{\hspace{0.2cm}\vec{\bf k}_{1}^{\,\perp}}{\circledast}
\hspace{-0.1cm}\Phi_{(2)}^{[\Gamma_1]}(\bar x_{2})\,,
\end{eqnarray}
where integration over $dk^-_1$ with $\delta((P_1-k_1)^2)$ has  been done and we introduce the notations
\begin{eqnarray}
\label{Conv-Delta}
&& F \stackrel{\hspace{0.2cm}\vec{\bf k}_{1}^{\,\perp}}{\circledast} G =
\int(d^2\vec{\bf k}_{1}^{\,\perp})  \, F(\vec{\bf k}_{1}^{\,\perp})
\\
&&
\times\int(d\ell^- d^2\vec{\ell}_{\,\perp}) \, \Delta(\ell^-,\vec{\ell}_\perp)\,
G(\vec{\bf k}_{1}^{\,\perp};\ell^- \vec{\ell}_{\,\perp})\,,
\nonumber\\
&& \Delta(\ell^-,\vec{\ell}_\perp)= \frac{1}{\ell^- - \vec{\ell}^{\,2}_\perp/(2x_{21}P_1^+) + i\,{\rm sign}(x_{21})\, 0}.
\end{eqnarray}

In Eqn.~(\ref{Lower-Phi-2}), the function $\mathbb{B}_{\alpha\beta}^{[\Gamma_2,\,\Gamma_1]}(x_1,x_2)$ (which is
expressed, generally speaking, through two non-collinear pion distribution amplitudes) is an analog of
the function $B^V(x_1,x_2)$ appeared in the inclusive channel \cite{Anikin:2010wz}.
The similarity of these two functions takes place provided the small invariant mass of spectators in the pion sector.
In other words, the mentioned similarity can be understood as the manifestation of duality between exclusive and inclusive channels.
Besides, if we have the restriction somewhat of
$\lvert \vec{\ell}_{\,\perp}\rvert\gg \lvert\vec{\bf k}_{1}^{\,\perp}\rvert$, the approximation can be implemented by two collinear distribution
amplitudes.

Let us discuss the pole at $x_1=x_2$ which is factorized as a prefactor in Eqn.~(\ref{Lower-Phi-2}).
This is exactly the gluon pole which has to be treated within the contour gauge frame as described
in \cite{Anikin:2010wz, Anikin:2015xka, Anikin:2015esa}, {\it i.e.}
\begin{eqnarray}
\label{GP}
\frac{1}{x_2-x_1} \stackrel{\text{c. g.}}{\Longrightarrow}
\frac{1}{x_2-x_1 - i\epsilon}.
\end{eqnarray}
Notice that the complex prescription emanates from the corresponding integral representation of
the theta function (see, \cite{Anikin:2015xka} for details).

Having performed the Lorentz decomposition in the corresponding matrix elements,
in the case we consider the $B^V$- function analog takes the form
\begin{eqnarray}
\label{Lower-Phi-3}
\mathbb{B}^{[\Gamma_2]}_{\alpha\beta}(x_1,x_2)=
\frac{1}{2}\,\frac{g^\perp_{\alpha\beta}\, V^{[\Gamma_2]}}{x_2-x_1 - i\epsilon}
\Phi_{(1)}^{\text{tw-3}}(x_1)\hspace{-0.2cm} \stackrel{\hspace{0.2cm}\vec{\bf k}_{1}^{\,\perp}}{\circledast}
\hspace{-0.1cm}\Phi_{(2)}^{\text{tw-2}}(x_2),
\end{eqnarray}
where the Lorentz tensor $V^{[\Gamma_2]}$ means $P^\perp_{1}$ or $\tilde n^{[\,-} P_1^{+\,]}$ for the pion-to-vacuum matrix elements and
$e^\perp$ for the rho-meson-to-vacuum matrix element. In Eqn.~(\ref{Lower-Phi-3}), $P_1^+$ in the numerator which originates from the parametrization
of $\Phi_{(2)}^{[\Gamma_1]}$ cancels the same component coming from the denominator of gluon propagator (see, Eqn.~(\ref{Lower-Phi-2})).

With these, after factorization the standard diagram hadron tensor reads
\begin{eqnarray}
\label{HadTen-St-2}
&&{\cal W}_{\mu\nu}^{(\text{stand.})}= i\, \int (dx_1)\,(dy) \, \delta^{(4)}(x_1P+yP_2-q)
\overline{\Phi}^{[\Gamma]}(y)
\nonumber\\
&&\times\int (dx_2)\,
\text{tr}\big[ \gamma_\nu \Gamma \gamma_\alpha \frac{x_{21}\hat P_1^+}{-x_{21} y s + i0}
\gamma_\mu \gamma^- \gamma_\beta \Gamma_2
\big]
\\
&&\times\frac{1}{2}\,\frac{g^\perp_{\alpha\beta}\, V^{[\Gamma_2]}}{x_2-x_1 - i\epsilon}
\Phi_{(1)}^{\text{tw-3}}(x_1)\hspace{-0.2cm} \stackrel{\hspace{0.2cm}\vec{\bf k}_{1}^{\,\perp}}{\circledast}
\hspace{-0.1cm}\Phi_{(2)}^{\text{tw-2}}(x_2)\,.
\nonumber
\end{eqnarray}
Here and below, for the sake of convenience, we single out the complex $i$ which emanates from either
the parametrization of upper blob, Eqn.~(\ref{Upper-Phi}), or from the parametrization of
pion-to-vacuum matrix element of twist two operator in the lower blob.
%


The next object of our discussion is the additional diagram, see Fig.~\ref{Fig-DY-1-2-b}, which contributes to
hadron tensor. According to \cite{Anikin:2010wz}, this is the so-called nonstandard diagram which
does not exist if $B^V$-function is real.

In principle, derivation of this part of hadron tensor is similar to what we present
for the standard diagram contribution. Before factorization, we have
\begin{eqnarray}
\label{HadTen-NonSt}
&&{\cal W}_{\mu\nu}^{(\text{nonstand.})}=\int (d^4k_1)\,(d^4k_2) \, \delta^{(4)}(k_1+k_2-q)
\nonumber\\
&&\times\int (d^4\ell)\, {\cal D}_{\alpha\beta}(\ell)
\text{tr}\big[ \gamma_\nu \Gamma \gamma_\mu S(k_1) \gamma_\alpha \Gamma_1 \gamma_\beta \Gamma_2
\big]
\nonumber\\
&&\times\bar\Phi^{[\Gamma]}(k_2)\,
\Phi_{(2)}^{[\Gamma_1]}(k_1; \ell)\,
\Phi_{(1)}^{[\Gamma_2]}(k_1)\, \delta\big( (P_1-k_1)^2 \big).
\end{eqnarray}
Then, we again perform the factorization procedure and, finally, the nonstandard diagram
hadron tensor is given by
\begin{eqnarray}
\label{HadTen-NonSt-2}
&&{\cal W}_{\mu\nu}^{(\text{nonstand.})}=i\,\int (dx_1)\,(dy) \, \delta^{(4)}(x_1P_1+yP_2-q)
\overline{\Phi}^{[\Gamma]}(y)
\nonumber\\
&&\times\,
\text{tr}\big[ \gamma_\nu \Gamma \gamma_\mu \frac{\gamma^+}{2x_1P_1^++i0} \gamma_\alpha \gamma^- \gamma_\beta \Gamma_2
\big] \frac{1}{2}\int (dx_2)
\\
&&\times\,\frac{g^\perp_{\alpha\beta}\, V^{[\Gamma_2]}}{x_2-x_1 - i\epsilon}
\Phi_{(1)}^{\text{tw-3}}(x_1)\hspace{-0.2cm} \stackrel{\hspace{0.2cm}\vec{\bf k}_{1}^{\,\perp}}{\circledast}
\hspace{-0.1cm}\Phi_{(2)}^{\text{tw-2}}(x_2)\,.
\nonumber
\end{eqnarray}


Last but not least, to get the gauge invariant expression for the hadron tensor
we sum the contributions of Eqns.~(\ref{HadTen-St-2}) and (\ref{HadTen-NonSt-2}). The sum reads
\begin{eqnarray}
\label{HadTen-GI}
&&\overline{\cal W}_{\mu\nu}= \int d^2\vec{\bf q}_\perp {\cal W}_{\mu\nu}
\\
&&=
i\,\int (dx_1)\,(dy) \, \delta(x_1P^+_1-q^+)\delta(yP^-_2-q^-)
\nonumber\\
&&\times  \bar F(y) \int (dx_2) \widetilde{B}(x_1,x_2) \,\frac{T^\nu}{P_1\cdot P_2}
\Big[ \frac{P^{\mu}_1}{y}- \frac{P^{\mu}_2}{x_1}\Big] \,,
\nonumber
\end{eqnarray}
where
\begin{eqnarray}
\bar F(y)=
          \begin{pmatrix}
          \bar f_T(y)\vspace{0.2cm}\\
          \bar h_{1}(y)
          \end{pmatrix}\quad
T^\nu =
          \begin{pmatrix}
          \varepsilon^{P_1+S_\perp P_2^\perp} V^\nu_{\perp}\vspace{0.2cm}\\
          \varepsilon^{\nu S_\perp P_2 P_1}_\perp
          \end{pmatrix}
\end{eqnarray}
and
\begin{eqnarray}
\label{B-fun}
\widetilde{B}(x_1,x_2)=\frac{1}{2}
\frac{\Phi_{(1)}^{\text{tw-3}}(x_1)\hspace{-0.2cm}
\stackrel{\hspace{0.2cm}\vec{\bf k}_{1}^{\,\perp}}{\circledast}
\hspace{-0.1cm}\Phi_{(2)}^{\text{tw-2}}(x_2)}{x_2-x_1 - i\epsilon}\,.
\end{eqnarray}
In Eqn.~(\ref{B-fun}), the pion distribution amplitudes of twist $2$ and $3$ include the different dimensionful
pre-factors.
Notice that the derived gauge invariant hadron tensor, see Eqn.~(\ref{HadTen-GI}), coincides formally with the results obtained in
\cite{Anikin:2010wz} for the usual Drell-Yan process.


We now calculate the single spin asymmetry, see Eqns.~(\ref{SSA-1}).
Within the CS-frame \cite{Arnold:2008kf, Barone:2001sp},
calculating the imaginary part of $\widetilde{B}(x_1,x_2)$ and
contracting the leptonic and hadron tensors, we obtain
(here, ${\cal L}_{\mu\nu}$ implies the unpolarized lepton tensor)
\begin{eqnarray}
\label{SSA}
&&{\cal L}_{\mu\nu}\, \Im\text{m}_B\,\overline{\cal W}_{\mu\nu} =
\bar F(y_B)\,
\Phi_{(1)}^{\text{tw-3}}(x_B)\hspace{-0.2cm} \stackrel{\hspace{0.2cm}\vec{\bf k}_{1}^{\,\perp}}{\circledast}
\hspace{-0.1cm}\Phi_{(2)}^{\text{tw-2}}(x_B)
\nonumber\\
&&\times
\frac{2}{x_B\, y_B} (\ell_1\cdot \hat z)\, \,
\begin{pmatrix}
(\ell^\perp_1\cdot V^{\perp}) \vec{\bf S}_\perp\wedge\vec{\bf P}_{2 \perp}\vspace{0.2cm}\\
\varepsilon^{\ell_1 S_\perp P_2 P_1}
\end{pmatrix}\,,
\end{eqnarray}
where
$(\ell_1\cdot \hat z)=-\frac{Q^2}{2}\cos\theta$,
$\varepsilon^{l_1 S^\perp P_1 P_2}
	= - \frac{s\, Q}{4} \,\lvert\vec{\bf S}_\perp\rvert \sin\theta \sin\phi_S$.

Therefore, for the chiral-odd contributions, we predict a new asymmetry which, in terms of \cite{COMPASS}, reads
\begin{eqnarray}
\label{NewSSA}
{\cal A}_T=\frac{S_{\perp}}{Q} \, \frac{  D_{[\sin 2\theta]} \, \sin\phi_S B^{\sin\phi_S}_{UT}}
{\bar f_1(y_B)\,H_1(x_B)},\, D_{[\sin 2\theta]}=\frac{\sin 2\theta}{1+\cos^2\theta}
\end{eqnarray}
where
$B^{\sin\phi_S}_{UT}=2\bar h_1(y_B)\, \Phi_{(1)}^{\text{tw-3}}(x_B)\hspace{-0.2cm} \stackrel{\hspace{0.2cm}\vec{\bf k}_{1}^{\,\perp}}{\circledast}
\hspace{-0.1cm}\Phi_{(2)}^{\text{tw-2}}(x_B)$;
$\bar f_1(y_B)$ and $H_1(x_B)$ stem from the unpolarized cross-section and they parameterize the following matrix elements:
\begin{eqnarray}
\label{Unpol}
&&\langle P_2| \bar\psi\, \gamma^- \,\psi |P_2 \rangle\stackrel{{\cal F}}{\sim} P^-_2\bar f_1(y),\,
\nonumber\\
&&\langle P_1| \bar\psi_+ |q(K)\rangle \langle q(K)| \psi_+ |P_1 \rangle\stackrel{{\cal F}}{\sim} P_1^+ H_1(x).
\end{eqnarray}
where
\begin{eqnarray}
\label{Unpol2}
&&H_1(x)=
\\
&&\frac{1}{2\bar x_1 P_1^+} \int(d^2\vec{\bf k}_{1}^{\,\perp})
\Phi^{[\gamma^+(\gamma_5)]}_{(2)}(\bar x_1,\vec{\bf k}_{1}^{\,\perp})
\Phi^{[\mathbb{I}(\gamma_5)]}_{(1)}(x_1,\vec{\bf k}_{1}^{\,\perp}).
\nonumber
\end{eqnarray}
Indeed, the leading twist Sivers asymmetry $A^{\sin\phi_S}_{UT}$, which formally stands at the similar tensor combination
$\varepsilon^{l^\perp_1 S^\perp P_1 P_2}$, appears only together with the depolarization factor $D_{[1+\cos^2\theta]}$.
In its turn, the higher twist asymmetries $A^{\sin(\phi_S\pm\phi)}_{UT}$ at $D_{[\sin 2\theta]}$
correspond to the different tensor structures, $\varepsilon^{q S^\perp P_1^\perp P_2}\sim \sin(\phi_S\pm\phi)$.


In conclusion, we derive the gauge invariant meson-induced DY hadron tensor with the essential
spin transversity and ``primordial'' transverse momenta. Our calculation includes both the
standard-like, which is well-known, and nonstandard-like diagram, which is first found in \cite{Anikin:2010wz}, contributions.
The latter contribution plays a crucial role for the gauge invariance.

In the paper, we focus on the case where one of pion distribution amplitudes has been projected onto the chiral-odd combination.
The latter singles out the chiral-odd parton distribution inside nucleons.
The chiral-odd tensor combinations are very relevant for the future experiments implemented by COMPASS \cite{COMPASS}.
We predict new single transverse spin asymmetries to be measured experimentally
which are associated with the spin transversity and with the
nontrivial $\varphi$-angular dependence.
The latter asymmetry can eventually be treated as a signal for the transverse ``primordial'' momentum dependence
which probes both gluon poles and time-odd functions.
In contrast to \cite{Brandenburg:1994wf}, our SSA is given by the
interference of two amplitudes rather then the square of a given amplitude.
We emphasize that the possibility to study different SSAs reachable in COMPASS experiments and
induced by chiral-odd and time-odd distribution functions/amplitudes appears only thanks for the gluon pole presence.
Thus, the proposed angular dependence of SSA
can give implicit evidences for the gluon pole observation in COMPASS experiments.

We model the analog of the collinear parton distribution function
$B^V(x_1,x_2)$-function of \cite{Anikin:2010wz} determined in inclusive channel
by means of the gluon propagator and two non-collinear exclusive meson distribution amplitudes of twist $2$ and $3$.
For this modelling, the invariant masses of undetected
spectators in the pion sector have to be considered as relatively small.
Our model demonstrates the manifestation of duality between different factorization regimes in exclusive
and inclusive channels (see, \cite{Anikin:2008bq}).

\vspace{0.5cm}

We thank A.V.~Efremov, L.~Motyka and A.~P.~Nagaytsev for useful discussions.
The work by I.V.A. was partially supported by the Bogoliubov-Infeld Program.
I.V.A. also thanks the Theoretical Physics Division of NCBJ (Warsaw) and
the Department of Physics at the University of Krakow for warm hospitality.
L.Sz. is partly supported by grant No $2015/17/\text{B}/\text{ST}2/01838$ from the National Science
Center in Poland.

\end{document}